# SU-8 meta phenylenediamine conjugated thin film for temperature sensing

**Hani Barhum[1,2,†,*], Muhammad A. Atrash[1,2], Inga Brice[3], Toms Salgals[4], Madhat Matar[2], Mariam Amer[1,2], Ziad Abdeen[5,6], Janis Alnis[3], Vjaceslavs Bobrovs[4,] Abdul Muhsen Abdeen[5,7], and Pavel Ginzburg[1]**

*[1]Department of Electrical Engineering, Tel Aviv University, Ramat Aviv, Tel Aviv 69978, Israel.*
*[2]Triangle Regional Research and Development Center, Kfar Qara' 3007500, Israel.*
*[3]Institute of Atomic Physics and Spectroscopy, University of Latvia, Jelgavas Street 3, 1004 Riga, Latvia.*
*[4]Institute of Telecommunications, Riga Technical University, 12 Azenes Street, 1048 Riga, Latvia.*
*[5]Al-Quds Public Health Society, Jerusalem, Palestine*
*[6]Al-Quds Nutrition and Health Research Institute, Al-Quds University, East Jerusalem, Palestine*
*[7]Marshall University John Marshall Dr, Huntington, WV 25755, United States*
*HB, https://orcid.org/0000-0003-0214-0288 ; MAA, https://orcid.org/0000-0002-2500-0811*

## Summary

Polymers, demonstrating distinctive optical properties alongside facile and mastered fabrication methods, have become increasingly important platforms for realizing a variety of nanophotonic devices. Enhancing these materials with additional functions might expand their range of multidisciplinary applications. Here, we demonstrate the temperature sensing potential of SU8-Phenylenediamine (SU8-mPD), which was produced by epoxy amination of the SU-8 polymer. The SU8-mPD properties were examined through a series of molecular structural techniques and optical methods. Thin layers have demonstrated optical emission and absorption in the visible range around 420 and 520 nm respectively alongside a strong thermal responsivity, characterized by the 18 ppm·K$^{-1}$ expansion coefficient. A photonic chip, comprising a thin 5-10 µm SU8-mPD layer, encased between parallel silver and/or gold thin film mirrors, has been fabricated. This assembly, when pumped by an external light source, generates a pronounced fluorescent signal which is superimposed with the Fabry-Pérot (FP) resonant response. The chip undergoes mechanical deformation in response to temperature changes, thereby shifting the FP resonance and encoding temperature information into the fluorescence output spectrum. The time response of the device was estimated to be below 500 msec opening a new avenue for optical sensing using SU8-based polymers. Thermoresponsive resonant structures, encompassing strong tunable fluorescent properties, can further enrich the functionalities of nanophotonic polymer-based platforms.

## Main Text

### Introduction

*Author for correspondence hani.barhom@gmail.com

†Present address: Department of Electrical Engineering, Tel Aviv University, Ramat Aviv, Tel Aviv 69978, Israel

Polymers have long been recognized for their versatility in diverse applications, ranging from structural materials to advanced electronic devices[1]. For example, polymer-based sensors are used to detect gas, humidity, pressure, specific molecules, and more[2–4]. Among a variety of available polymers, SU-8 has a special nische[5–7]. SU-8 is a negative, epoxy-type, near-UV photoresist derived from the EPON SU-8 epoxy resin[8]. Initially developed for realizing lithographic masks for microelectronics, it became a platform for high-aspect-ratio microstructures[9]. Beyond microelectronics, SU-8 found its niche in microfluidics[10–13], MEMS (Micro-Electro-Mechanical Systems), and biomedicine, showing adaptability across various domains[14]. SU-8 also has remarkable optical properties. It is inherently transparent in the visible spectrum and has a refractive index higher than glass, these, combined with its well-established methodology in fabrication, inspired a variety of optical applications[15,16] including waveguides and resonators[5]. Considering the remarkable optical properties alongside the facile fabrication approaches to treat SU-8, this material has the potential to serve as a multifunctional platform in nanophotonic applications [17]. One among the variety of options is to explore the temperature capabilities of SU-8-based devices. Given the strong temperature-dependent characteristics of SU-8, it can be considered an excellent candidate for such applications[18–20]. Furthermore, the compatibility of SU-8 with microfabrication techniques allows for the development of miniaturized sensors that can be integrated into complex systems, offering real-time temperature monitoring with high spatial resolution. For example, SU-8 was used in the thermal flow sensor[21].

Optical methods for temperature sensing play a role in monitoring industrial processes, environment, and medical diagnostics to name a few[22–24]. Among a variety of possible architectures, interferometric methods are among the most accurate. Recently reported temperature sensors based on optical fiber architectures, especially those utilizing the Fabry-Pérot (FP) effect, have demonstrated superior sensitivities, response time, and temperature resolution. For example, 84.6 pm°C[-1] sensitivity alongside a fast response time of 0.51 ms, has been demonstrated[18]. A miniaturized optical fiber tip sensor for high-temperature applications with a sensitivity of 0.01226 nm°C[-1], suitable for environments up to 1000 °C has been shown[25]. Another high-temperature sensor based on a sapphire fiber FP interferometer, capable of withstanding temperatures as high as 1550 °C with a sensitivity of 32.5 pm°C[-1] [26] was reported. These studies exemplify the versatility and adaptability of FP sensor technologies across a wide range of temperature conditions. The essence of FP sensor operation is the shift of the optical resonance with refractive index changes. Furthermore, variations in FP sensor designs further highlight their possible perspective potential [27,28].

Remarkably, the optical characteristics of SU-8, and particularly its refractive index, are susceptible to temperature changes. Therefore, it is a prime candidate for use in optical temperature sensors[29].

Here we explore thermoresponsive properties of SU8-meta Phenylenediamine (SU8-mPD)[30–32], synthesized through epoxide amination of SU-8 to impart fluorescence abilities to the film. The optical properties, chemical structure, and photoluminescence of this material are investigated at different temperatures, and key parameters are extracted. Based on those results, we demonstrate a chip consisting of a thin SU8-mPD layer, sandwiched between two parallel metal mirrors - silver and gold thin films. This structure is optically pumped by an external light source and, as a result, a strong fluorescent signal is generated. The device's output signal is the fluorescent signal, superimposed with FP resonant response. SU8-mPD layer, experiencing mechanical deformation in response to the temperature change, changes FP resonant conditions and, as a result, the output fluorescent spectrum imprints the information on the temperature – Figure 1. The fluorescent-based architecture has an advantage over the passive resonant FP realization as it potentially relaxes the constraints on optical system alignment and thus can be advantageous in remote sensing scenarios.

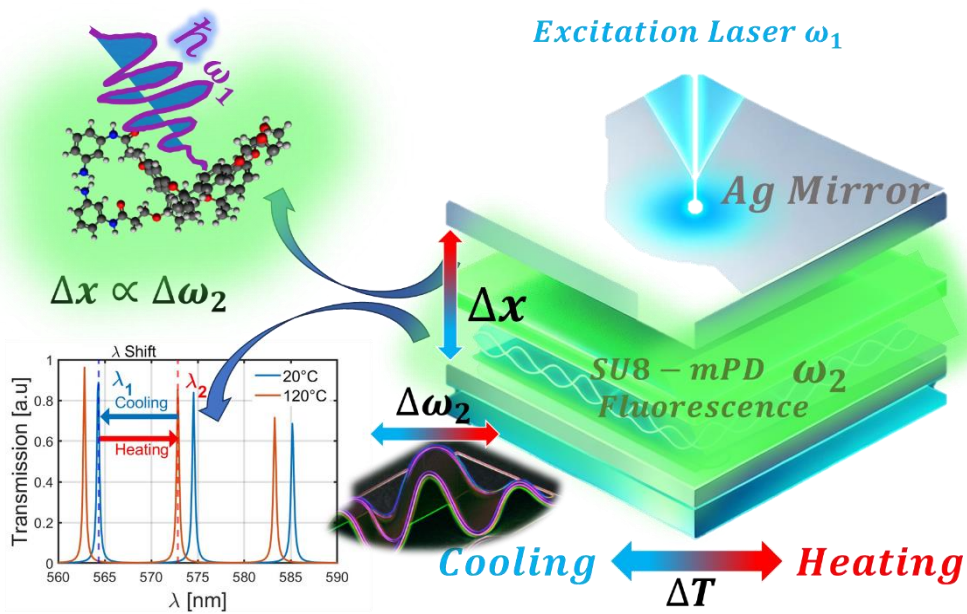

*Figure 1: Schematics of FP-based temperature sensor, encompassing two mirrors and a thermoresponsive conjugated fluorescent polymer. The structure is pumped at $\omega_1$ and emits light at a temperature-dependent wavelength. The measurement is the shift of the emission peak, i.e., '$\Delta\omega_2$', which depends on the distance between the mirrors. Thermoresponsive polymer experiences contraction or expansion '$\Delta x$', depending on the temperature change, which is monitored by '$\Delta\omega_2$'.*

## Methods

### Materials

All materials were procured from the following sources: photoresist 3005, 3050, methyl ether acrylate (PEGMEA), and SU8 developer were obtained from Bachem (UK) Ltd. Meta-phenylenediamine (m-PD), and Hydrochloric Acid (HCl) were sourced from Merck Ltd. All experiments utilized 18MΩ deionized water. ITO Slides 100nm on BK7 glass from Sigma Aldrich.

### SU8 Conjugation to Phenylenediamine

The desired meta phenylenediamine isomer (m-PD) was dissolved in either SU8 developer or PEGMEA through sonication. This saturated solution was then added in varying quantities to a 5 mL polymer solution and mixed thoroughly. The color transition of the SU8 solution from light yellow to dark green was observed after 24 hours, indicating that spectral changes stopped, and the reaction ended. HCl from a concentrated solution was added to the reaction to tune 5mM.

### Thin Film Deposition

Clean ITO-covered glass microscopes, after being diced into 2x2cm squares, were used in a clean room atmosphere. The metallic mirrors were evaporated on top of the ITO and SU8-mPD layers by e-beam evaporation. VST evaporator was technically used at $10^{-7}$ atm. The evaporation was done at slow rates of $0.4AS^{-1}$. After finalizing the process, the samples were stable for two years.

### Optical Properties of Conjugated SU8

Optical properties were determined using photoluminescence excitation (PLE) spectroscopy with a Synergy H1 plate reader. Absorbance spectra were measured using a Macys1100 spectrophotometer equipped with a tungsten lamp source and a silicon photodiode detector. Before testing, samples were diluted with deionized water. Fluorescence lifetime measurements were conducted using a PicoQuant system with a Taiko picosecond diode as the 375 nm excitation source. Confocal images were obtained using the Leica 8 system. The materials were deposited using an ion beam evaporator, with silver (Ag) and SU8 being spin-coated. Transmission and reflection spectra were obtained using a custom-made setup with an Andor spectrometer. Temperature control during heating and cooling experiments was achieved using a dedicated apparatus, and some experiments were conducted in a vacuum chamber equipped with a temperature controller.

### ATR-Fourier Transform Infrared Spectroscopy (FTIR) and Proton Nuclear Magnetic Resonance ([1]H-NMR)

A Nicolet iS10 FTIR Spectrometer equipped with a KBr/Ge beam splitter and a DTGS detector was used to take FTIR spectra. Dry samples were analyzed by ASTM E1421 standards. For the [1]H-NMR analysis, dried and purified SU8-MPD powder was dissolved in chloroform and analyzed using a Bruker Ascend 500 High-Resolution NMR machine.

**Atomic Force and Electron Scanning Microscopy**

Film surface and cross-sectional thickness were examined using an environmental electron scanning microscope (ESEM). Surface roughness was further characterized using a JPLEC atomic force microscope (AFM).

## Results and discussion

The FP-based temperature sensor is based on a thermoresponsive polymer (SU8-meta Phenylenediamine – SU8-mPD) layer, sandwiched between two metalized mirrors.

The mirrors are formed by a BK-7 glass slide, coated with 100 nm indium tin oxide (ITO) and 30 nm metal via ion beam evaporation. The aminated epoxy is then spin-coated on top of the processed BK-7 glass slide. After the soft baking process, another 30 nm layer of metal is deposited to form the upper mirror. Both silver (Ag) and gold (Au) coatings were assessed in this study. Fabrication details are elaborated in the *Methods section.*

*Film characterization*

In the first step, several characterization techniques were applied to reveal the layer parameters and surface quality of the device. Scanning electron microscope (SEM) images of the edge were acquired - Figure 2(a-b).

While analyzing SEM images enables extracting information on actual layer thickness, confocal imaging allows assessing the uniformity of mPD distribution inside the polymer. The layer thickness as observed in Figure 2(b) is nearly 10 μm. Figure 2(c) demonstrates a nearly homogeneous light-emitting layer of ~12 μm thickness (pump at 488 nm, emission at 505-700nm band, detailed investigation of optical properties will follow). The fluorescent layer thickness was estimated by fitting the cross-sectional intensity with a Gaussian profile, as it appears in Figure 2(c). Atomic Force Microscopy (AFM) was used to probe the surface roughness of the mirrors, which might be affected by ion beam evaporation and inhomogeneity of SU8 in the case of the upper mirror. Figure 2(d-e) demonstrates the colormap and the corresponding 3D view of the surface, indicating the topology variations below 3 nm, maintained over large-scale areas, prevailing microns across. To investigate the topology variations on the millimeter to centimeter scale, the device was imaged after illuminating the structure with a 532 nm LED source and obtaining an interference pattern from SU8-mPD (optical details will follow) - Figure 2(f). Long-range variations of contrast demonstrate micron scale (wavelength comparable to creating an interference speckle) imperfections within the film. Nevertheless, finding homogeneous areas with sizes, prevailing the focal spot size of the collection objective is rather trivial and will be done in the subsequent realization of the temperature sensor.

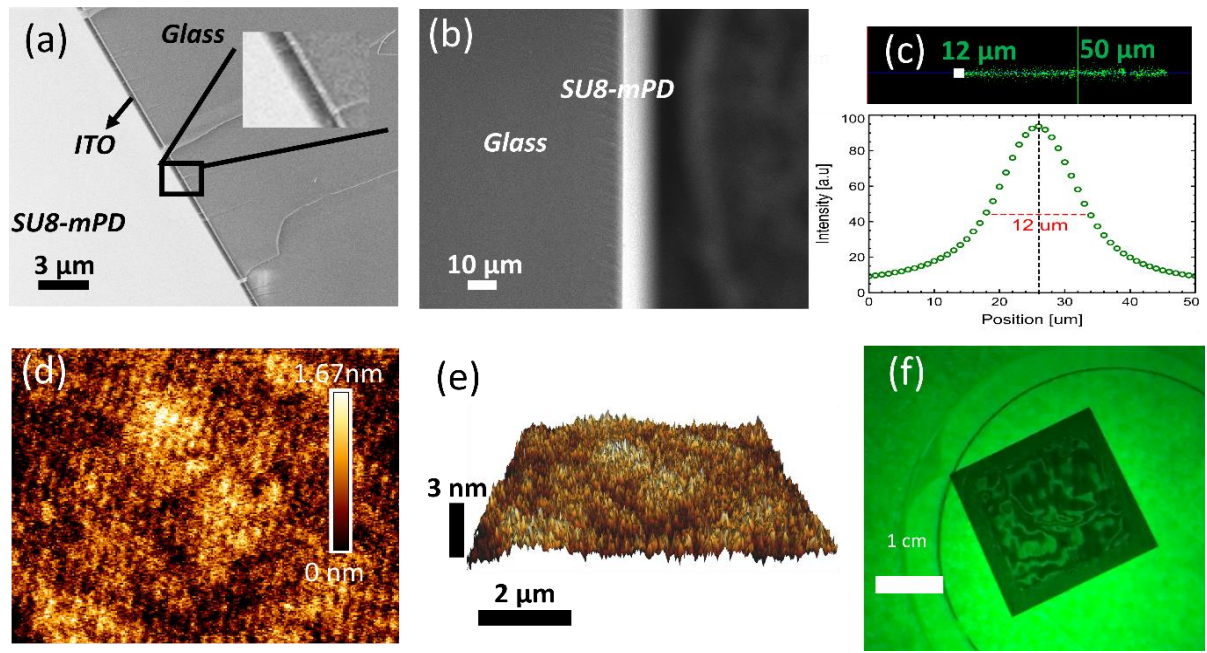

*Figure 2: Characterization of SU8-mPD conjugated thin films. (a)-(b) Cross-section SEM images of SU8 films on a glass coverslip, (c) Cross-section of the fluorescent film as observed in the confocal microscope (488 nm pump). The fluorescent profile is fitter with Gaussian and the red arrow indicates its 12 μm width. (d)-(e) High-resolution AFM image of the surface roughness in 2D and 3D layouts, respectively. (f) Image of the chip under 532nm LED showing interference patterns.*

*Optical and Structural Properties*

The optical properties of the structure, including absorption, fluorescence, and photoluminescence excitation (PLE), are systematically examined next and summarized in Figure 3(a). The absorption and PLE peaks of the SU8-mPD film are centered at 420 nm, which are not found in pure SU8 thus indicating the impact of the conjugation. PLE is observed at 530 nm, where the emission spectrum also has a maximum. Notably, unmodified SU8 emits at 450 nm under UV excitation, but this emission disappears after the polymer is crosslinked. Additionally, pure mPD emission peaked around ~440nm (violet-blue) when excited with UV light[33]. Therefore, the observed green emission is likely a result of the cross-linking of the aminated epoxy terminals with mPD.

To explore the chemical bonding within the polymer, a Fourier-transform infrared spectroscopy (FTIR) was applied - Figure 3(b). The obtained spectrum reveals several peaks, indicating relevance to the structure bonds. The peak at 1055 cm$^{-1}$ correlates to C-N stretching vibrations, signifying the presence of secondary amine bonds, which are known to contribute to green emissions. The peak at 3700 cm$^{-1}$ indicates hydroxyl groups, suggesting epoxy ring opening, while the 2900 cm$^{-1}$ peak relates to C-H components [34]. This data provides insights into the amination process of SU8.

Further chemical characterization was conducted using $^1$H-NMR, as shown in Figure 3(c). This analysis identifies three main regions: 1-2 ppm (aliphatic protons resonance), 2.5-4.5 ppm (secondary or tertiary amine structures), and 6.5-7.5 ppm (aromatic protons). A comparison with native SU8 revealed an increase in aromatic and aliphatic bonds, likely due to mPD integration. The proposed polymer structure, included as an inset in Figure 3(c), is based on these observations. XPS analysis, Figure 3(d-f), further elucidated the material's chemical composition and electronic state. The presence of C, N, and O were detected, with binding energies of 285, 399, and 533 eV, respectively, and corresponding atomic weights of 76%, 1.3%, and 22.7%. The C(1s) peak was separated into two components: C-C/C-H at 284.7 eV and C-O/C-N at 286.4 eV. Both N(1s) and O(1s) peaks were resolved to single chemical bonds. Furthermore, the evidence of these structural changes suggests the amination of the epoxy chain, leading to the green emission band. Notably, this green emission from the amines did not significantly alter the polymer's mechanical properties or its lithographic processability. As a result, SU8 maintains its inherent properties while exhibiting a new optical response.

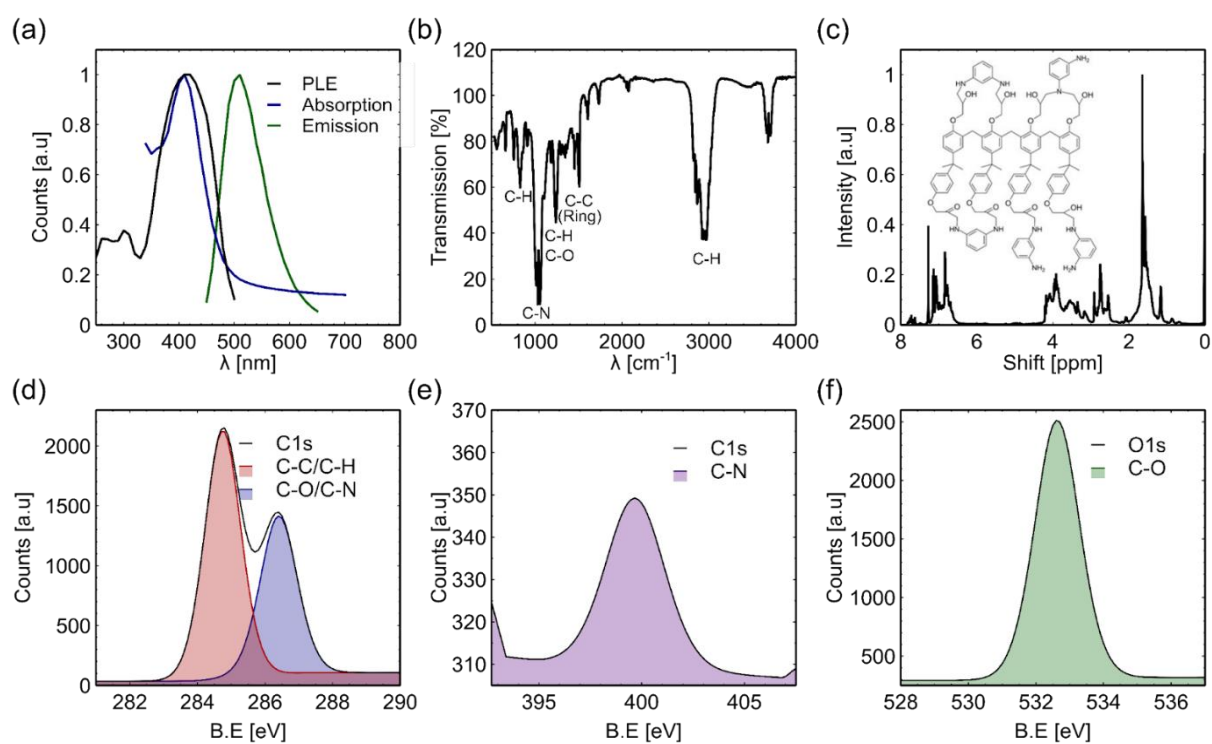

*Figure 3: Optical and structural properties of SU8-mPd layer. (a) PLE (observed at 530 nm), absorption, and emission (excitation at 420nm). (b) FTIR spectrum reveals the presence of C-H, C-C, C-N, and C-O bonds. (c) NMR shift - three regions at 1.6, 3.5, and 7 ppm. The suggested polymer structure appears as an inset. (d)-(f) XPS data show the presence of carbon(C(1s)), nitrogen (N(1s)) and oxygen (O(1s)), with an atomic weight of 76 %, 1.3 %, and 22.7 %, respectively.*

*Thermoresponsive Properties*

Given the refractive indexes of the materials are known, the thickness of the layers can be assessed from Free Spectral Range (FSR) measurements, performed by illuminating the FP cavity with white light (Supplementary Information, S1). Given the FSR value of approximately 9.8 THz and the refractive index of 1.56, the cavity length was calculated to be around 10 micrometers (Figure S1) in full correspondence with SEM and confocal analysis (Figure 2).

The same technique can be applied to monitor the thermal expansion mechanism and retrieve the relevant thermoresponsive coefficients. For this purpose, the temperature was linearly increased from room temperature (23.5°C) to 100°C, with measurements taken every 500 msec – results are summarized in the colormap (Figure 4). This time (500msec) was found to ensure the adiabatic expansion and will be then referred to as the *response time* of the detector. Considering the FSR formula, the thermal expansion coefficient is estimated as 18 ppm·K⁻¹, and the linear behavior in the range of the considered parameters is justified. The data is well corresponded with other reports[35]. Notably, the sample demonstrated remarkable stability, maintaining its properties without degradation for two years, during which the experiments were performed.

Figure 4(a) is the colormap, demonstrating the white light transmission at different temperatures. Figure 4(b) is the peak location as the function of the temperature. The linear approximation has been applied to the peak, marked with a thin solid black line on panel (a). The thermal expansion coefficient, estimated as 18 ppm·K⁻¹, comes from this fit.

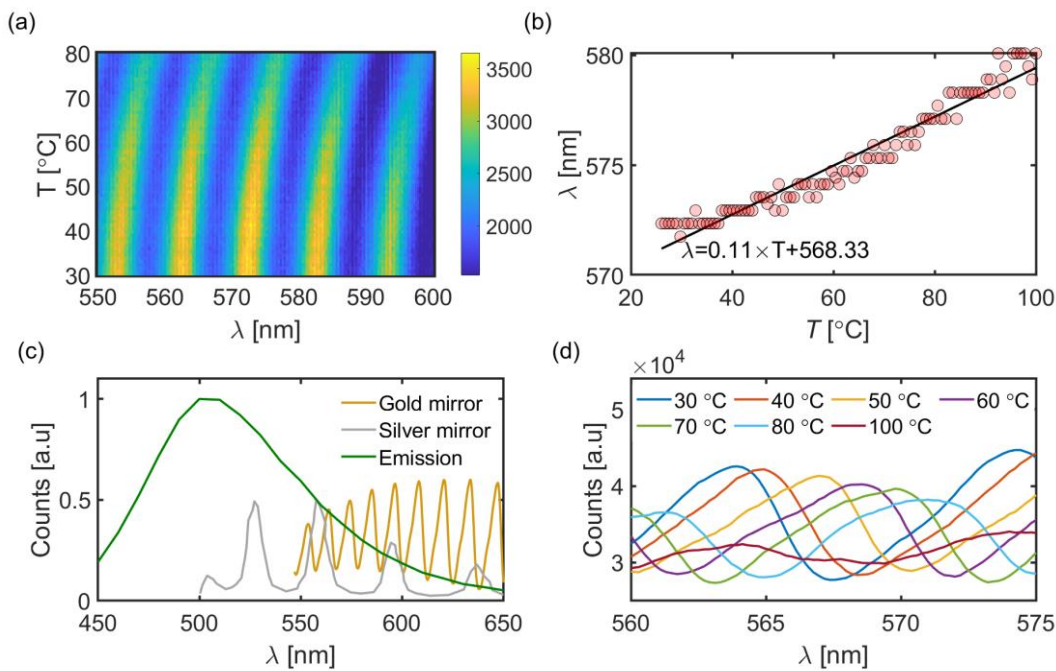

*Figure 4: (a) Temperature-dependent evolution of the FP-SU8-mPD transmission. (b) Transmission peak evolution with temperature. The analysis is done on the branch, indicated with a thin black line in panel (a). (c) Green curve - emission spectrum of SU8-mPD, yellow – 10 µm-thick FP with gold mirrors transmission, and gray - 5 µm-thick FP with silver mirrors transmission. (d) Zoomed peak shift of transmission spectra at different temperatures (gold FP).*

*Temperature Sensing*

To explore thermal sensitivity, photoluminescence was recorded at a span of temperatures during heating and cooling cycles. The layered structure (with mirrors) was exited at 420 nm and the emission spectrum was acquired. The temperature changes were generated with a closed-loop thermo-electric cooler, attached to the sample. Adiabatic changes were induced to ensure obtaining a uniform steady-state temperature along the sample (recall the 500 msec response time from the previous section). The fluorescent-based temperature detection might have practical advantages over simple white light monitoring in remote sensing applications, mitigating a need to align the incident light with the device orientation.

Figure S3(a, b) demonstrates the colormaps, attributed to the heating and cooling cycles. Each horizontal line across the map is the emission spectrum at a certain temperature. The periodic behavior of peaks is the result of the broad fluorescent signal (Figure 3(a)), modulated with the FP, acting as a filter. Numerical analysis, considering superimposing the fluorescent signal with the FP cavity, acting as a filter, supports the experimentally observed data (Supplementary, S2). The experimentally obtained colormaps demonstrate redshifts to higher wavelengths at elevated temperatures due to the mechanical expansion of the SU8 layer, increasing the distance between the metallic mirror layers. Conversely, during cooling, a blue shift to lower wavelengths occurred due to contraction and resonator shrinkage. Additionally, a decrease in photoluminescence intensity at higher temperatures comes from accelerating nonradiative recombination channels.

*Sensitivity assessment*

The resonance wavelengths as a function of temperature for both heating (red) and cooling (blue) are extracted and appear in Figure S3(c). We observed that the resonances shifted as the SU8-mPD layer thermally expanded and shrunk. Figures 5(a) and (b), extracted from the heating cycle (data from Figure S3), show the photoluminescence intensity for several environmental temperatures, experienced by the devices with gold and silver mirrors. Note, that the SU8-mPD layer thickness is different – 10 and 5μm, respectively. The evolution of the peak position, initially located at 620nm gold and 556nm silver (both at room temperature), is followed and fitted with a linear law. Figures 5(c) and (d) show the results and the extracted slope - 0.16 and 0.12 nm∘C$^{-1}$, respectively. The maximum intensity of photoluminescence as a function of temperature is shown in Figure 5 (c, inset) with a fit to first- and second-order polynomials. The best fit was found with a second-order polynomial. The latter showed another frame of reference to quantify the changes related to temperature monitoring. Worth noting that the absolute values of those coefficients have no physical meaning at the light intensity depending on the collection optics. However, what does have quantitative measures is the ratio between the fitting coefficients in the case of the linear fit and a more complex low in the case of the quadratic.

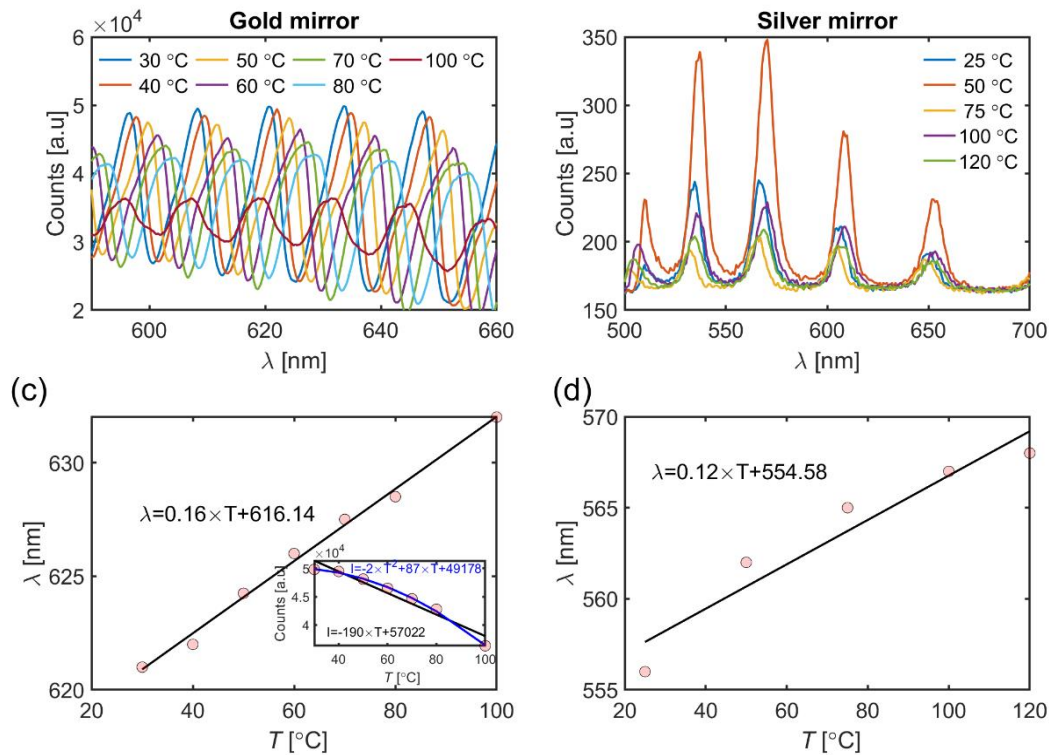

*Figure 5. Photoluminescent intensity [a.u.] at different temperatures for Fabry-Perot devices with (a) gold and (b) silver mirrors. SU8-mPD layer thicknesses are 10μm and 5μm, respectively. (c) Resonance wavelength as a function of temperature (for gold mirror) and the corresponding linear fit. Inset - photoluminescence intensity (of the peak) as a function of temperature and the first and second-order polynomial fits. (d) Resonance wavelength as a function of temperature for silver mirror device.*

After extracting the sensitivity parameter, it can be assessed versus other recent realizations of temperature sensors, based on optical techniques, which are surveyed in Table 1. The reported realization has comparable performance and dynamics range. Furthermore, both parameters can be improved by, e.g., making a weighted average between the peak positioning. Dynamic range is potentially much larger than reported, nevertheless, it was not confirmed with the experiment owing to the equipment limitations.

## Conclusions

The temperature-responsive characteristics of SU8-mPD, synthesized through epoxide amination of SU8, using a range of tools including XPS, [1]H-NMR, AFM, and SEM were explored. These techniques provided insights into the material's chemical and optical properties, relating them to the thermal-induced deformation of the material. Subsequently, thin SU8-mPD layers were encased between two metal mirrors to create a Farby-Perot cavity, susceptible to environmental temperature changes. A series of experiments, based on white-light transmission and fluorescent emission revealed the sensitivity of the device to the temperature variations. The operational principle of the chip is the superposition of the fluorescent signal with Fabry-Perot response, which strongly depends on the distance between the mirrors, which is governed by the thermal expansion of the polymer layer. This configuration has several advantages over a straightforward white light transmission, as it relaxes several tight requirements on arranging the optical system, including a demand for an accurate alignment and coherence of the expiation source. The demonstrated performance of the chip, i.e., 180 pm/°C, extended dynamic range, and 500 msec response time, puts it in line with other recent optical sensors and tend to outperform them in the ease of fabrication, cost, and handling. Functional SU8-based materials open new applied avenues across many disciplines, including nanophotonics, microfluidics, environmental monitoring, biomedical diagnostic chips, and many others.

## Acknowledgments

ERC StG "InMotion" (802279), QuanTAU - Center for Quantum Science and Technology equipment grant, RTU Team acknowledges the Latvian Council of Science project: No. lzp-2022/1-0553.

# References

(1)     Yuan, X.; Liu, H.; Sun, B. N-Doped Carbon Dots Derived from Covalent Organic Frameworks Embedded in Molecularly Imprinted Polymers for Optosensing of Flonicamid. *Microchemical Journal* **2020**, *159*. https://doi.org/10.1016/j.microc.2020.105585.

(2)     Cichosz, S.; Masek, A.; Zaborski, M. Polymer-Based Sensors: A Review. *Polymer testing* **2018**, *67*, 342–348. https://doi.org/10.1016/j.polymertesting.2018.03.024.

(3)     Uchiyama, S.; Matsumura, Y.; de Silva, A. P.; Iwai, K. Fluorescent Molecular Thermometers Based on Polymers Showing Temperature-Induced Phase Transitions and Labeled with Polarity-Responsive Benzofurazans. *Analytical chemistry* **2003**, *75* (21), 5926–5935. https://doi.org/10.1021/ac0346914.

(4)     McQuade, D. T.; Pullen, A. E.; Swager, T. M. Conjugated Polymer-Based Chemical Sensors. *Chemical reviews* **2000**, *100* (7), 2537–2574. https://doi.org/10.1021/cr9801014.

(5)     LaBianca, N. C.; Gelorme, J. D. High-Aspect-Ratio Resist for Thick-Film Applications; 1995; Vol. 2438. https://doi.org/10.1117/12.210413.

(6)     Bêche, B.; Gaviot, E.; Godet, C.; Zebda, A.; Potel, A.; Barbe, J.; Camberlein, L.; Vié, V.; Panizza, P.; Loas, G.; Hamel, C.; Zyss, J.; Huby, N. Spin Coating and Plasma Process for 2.5D and Hybrid 3D Micro-Resonators on Multilayer Polymers; 2009; Vol. 7356. https://doi.org/10.1117/12.820339.

(7)     Bêche, B. Integrated Photonics Devices on SU8 Organic Materials. *International Journal of Physical Sciences* **2010**, *5* (6).

(8)     J Zhang; K L Tan; G D Hong; L J Yang; H Q Gong. Polymerization Optimization of SU-8 Photoresist and Its Applications in Microfluidic Systems and MEMS. *Journal of Micromechanics and Microengineering* **2001**, *11* (1), 20. https://doi.org/10.1088/0960-1317/11/1/304.

(9)     Liu, J.; Cai, B.; Zhu, J.; Ding, G.; Zhao, X.; Yang, C.; Chen, D. Process Research of High Aspect Ratio Microstructure Using SU-8 Resist. *Microsystem Technologies* **2004**, *10* (4), 265–268. https://doi.org/10.1007/s00542-002-0242-2.

(10)   Lin, C. H.; Lee, G. B.; Chang, B. W.; Chang, G. L. A New Fabrication Process for Ultra-Thick Microfluidic Microstructures Utilizing SU-8 Photoresist. *Journal of Micromechanics and Microengineering* **2002**, *12* (5). https://doi.org/10.1088/0960-1317/12/5/312.

(11) Hamdi, F. S.; Woytasik, M.; Couty, M.; Francais, O.; Le Pioufle, B.; Dufour-Gergam, E. Low Temperature Irreversible Poly(DiMethyl) Siloxane Packaging of Silanized SU8 Microchannels: Characterization and Lab-on-Chip Application. *Journal of Microelectromechanical Systems* **2014**, *23* (5). https://doi.org/10.1109/JMEMS.2014.2331454.

(12) Talebi, M.; Cobry, K.; Sengupta, A.; Woias, P. Transparent Glass/SU8-Based Microfluidic Device with on-Channel Electrical Sensors; 2017. https://doi.org/10.3390/proceedings1040336.

(13) Talebi, M.; Woias, P.; Cobry, K. Analysis of Impedance Data from Bubble Flow in a Glass/SU8 Microfluidic Device with on-Channel Sensors. *Sensors and Actuators, A: Physical* **2018**, *279*. https://doi.org/10.1016/j.sna.2018.07.004.

(14) Ransley, J. H. T.; Watari, M.; Sukumaran, D.; McKendry, R. A.; Seshia, A. A. SU8 Bio-Chemical Sensor Microarrays. *Microelectronic Engineering* **2006**, *83* (4-9 SPEC. ISS.). https://doi.org/10.1016/j.mee.2006.01.175.

(15) Uchiyamada, K.; Okubo, K.; Yokokawa, M.; Carlen, E. T.; Asakawa, K.; Suzuki, H. Micron Scale Directional Coupler as a Transducer for Biochemical Sensing. *Optics Express* **2015**, *23* (13), 17156–17168. https://doi.org/10.1364/OE.23.017156.

(16) Su, Y.; Liu, C.; Brittman, S.; Tang, J.; Fu, A.; Kornienko, N.; Kong, Q.; Yang, P. Single-Nanowire Photoelectrochemistry. *Nature nanotechnology* **2016**, *11* (7), 609–612. https://doi.org/10.1038/nnano.2016.30.

(17) Golvari, P.; Kuebler, S. M. Fabrication of Functional Microdevices in SU-8 by Multi-Photon Lithography. *Micromachines* **2021**, *12* (5), 472. https://doi.org/10.3390/mi12050472.

(18) Liu, G.; Han, M.; Hou, W. High-Resolution and Fast-Response Fiber-Optic Temperature Sensor Using Silicon Fabry-Pérot Cavity. *Optics express* **2015**, *23* (6), 7237–7247. https://doi.org/10.1364/OE.23.007237.

(19) Zhang, J.; Sun, H.; Rong, Q.; Ma, Y.; Liang, L.; Xu, Q.; Zhao, P.; Feng, Z.; Hu, M.; Qiao, X. High-Temperature Sensor Using a Fabry-Perot Interferometer Based on Solid-Core Photonic Crystal Fiber. *Chinese Optics Letters* **2012**, *10* (7), 070607–070607.

(20) Zhang, L.; Jiang, Y.; Gao, H.; Jia, J.; Cui, Y.; Wang, S.; Hu, J. Simultaneous Measurements of Temperature and Pressure with a Dual-Cavity Fabry–Perot Sensor. *IEEE Photonics Technology Letters* **2018**, *31* (1), 106–109. https://doi.org/10.1109/LPT.2018.2885337.

(21) Vilares, R.; Hunter, C.; Ugarte, I.; Aranburu, I.; Berganzo, J.; Elizalde, J.; Fernandez, L. J. Fabrication and Testing of a SU-8 Thermal Flow Sensor. *Sensors and Actuators B: Chemical* **2010**, *147* (2), 411–417. https://doi.org/10.1016/j.snb.2010.03.054.

(22) Ho, C. K.; Robinson, A.; Miller, D. R.; Davis, M. J. Overview of Sensors and Needs for Environmental Monitoring. *Sensors* **2005**, *5* (1), 4–37. https://doi.org/10.3390/s5010004.

(23) Dincer, C.; Bruch, R.; Costa-Rama, E.; Fernández-Abedul, M. T.; Merkoçi, A.; Manz, A.; Urban, G. A.; Güder, F. Disposable Sensors in Diagnostics, Food, and Environmental Monitoring. *Advanced Materials* **2019**, *31* (30), 1806739. https://doi.org/10.1002/adma.201806739.

(24) Chen, X.; Leishman, M.; Bagnall, D.; Nasiri, N. Nanostructured Gas Sensors: From Air Quality and Environmental Monitoring to Healthcare and Medical Applications. *Nanomaterials* **2021**, *11* (8), 1927. https://doi.org/10.3390/nano11081927.

(25) Zhu, C.; Zhuang, Y.; Zhang, B.; Muhammad, R.; Wang, P. P.; Huang, J. A Miniaturized Optical Fiber Tip High-Temperature Sensor Based on Concave-Shaped Fabry–Perot Cavity. *IEEE Photonics Technology Letters* **2018**, *31* (1), 35–38. https://doi.org/10.1109/LPT.2018.2881721.

(26) Wang, B.; Niu, Y.; Zheng, S.; Yin, Y.; Ding, M. A High Temperature Sensor Based on Sapphire Fiber Fabry-Perot Interferometer. *IEEE Photonics Technology Letters* **2019**, *32* (2), 89–92. https://doi.org/10.1109/LPT.2019.2957917.

(27) Li, J.; Li, Z.; Yang, J.; Zhang, Y.; Ren, C. Microfiber Fabry-Perot Interferometer Used as a Temperature Sensor and an Optical Modulator. *Optics & Laser Technology* **2020**, *129*, 106296. https://doi.org/10.1016/j.optlastec.2020.106296.

(28) Zhao, L.; Zhang, Y.; Chen, Y.; Wang, J. Composite Cavity Fiber Tip Fabry-Perot Interferometer for High Temperature Sensing. *Optical Fiber Technology* **2019**, *50*, 31–35. https://doi.org/10.1016/j.yofte.2019.01.027.

(29)  Li, M.; Liu, Y.; Qu, S.; Li, Y. Fiber-Optic Sensor Tip for Measuring Temperature and Liquid Refractive Index. *Optical Engineering* **2014**, *53* (11), 116110–116110. https://doi.org/10.1117/1.OE.53.11.116110.

(30)  Barhum, H.; Kolchanov, D. S.; Attrash, M.; Unis, R.; Alnis, J.; Salgals, T.; Yehia, I.; Ginzburg, P. Thin-Film Conformal Fluorescent SU8-Phenylenediamine. *Nanoscale* **2023**. https://doi.org/10.1039/D3NR02744A.

(31)  Fatkhutdinova, L. I.; Barhum, H.; Gerasimova, E. N.; Attrash, M.; Kolchanov, D. S.; Vazhenin, I. I.; Timin, A. S.; Ginzburg, P.; Zyuzin, M. V. Metal Ion Sensing with Phenylenediamine Quantum Dots in Blood Serum. *ACS Applied Nano Materials* **2023**. https://doi.org/10.1021/acsanm.3c04494.

(32)  Barhum, H.; Alon, T.; Attrash, M.; Machnev, A.; Shishkin, I.; Ginzburg, P. Multicolor Phenylenediamine Carbon Dots for Metal-Ion Detection with Picomolar Sensitivity. *ACS Applied Nano Materials* **2021**, *4* (9), 9919–9931. https://doi.org/10.1021/acsanm.1c02496.

(33)  Do, M. T.; Nguyen, T. T. N.; Li, Q.; Benisty, H.; Ledoux-Rak, I.; Lai, N. D. Submicrometer 3D Structures Fabrication Enabled by One-Photon Absorption Direct Laser Writing. *Optics express* **2013**, *21* (18), 20964–20973. https://doi.org/10.1364/OE.21.020964.

(34)  Barhum, H.; McDonnell, C.; Alon, T.; Hammad, R.; Attrash, M.; Ellenbogen, T.; Ginzburg, P. Organic Kainate Single Crystals for Second-Harmonic and Broadband THz Generation. *ACS Applied Materials and Interfaces* **2022**. https://doi.org/10.1021/acsami.2c18454.

(35)  Feng, R.; Farris, R. J. The Characterization of Thermal and Elastic Constants for an Epoxy Photoresist SU8 Coating. *Journal of materials science* **2002**, *37*, 4793–4799. https://doi.org/10.1023/A:1020862129948.

(36)  Gomes, A. D.; Becker, M.; Dellith, J.; Zibaii, M. I.; Latifi, H.; Rothhardt, M.; Bartelt, H.; Frazão, O. Multimode Fabry–Perot Interferometer Probe Based on Vernier Effect for Enhanced Temperature Sensing. *Sensors* **2019**, *19* (3), 453. https://doi.org/10.3390/s19030453.

(37)  Liao, C. R.; Hu, T. Y.; Wang, D. N. Optical Fiber Fabry-Perot Interferometer Cavity Fabricated by Femtosecond Laser Micromachining and Fusion Splicing for Refractive Index Sensing. *Optics express* **2012**, *20* (20), 22813–22818. https://doi.org/10.1364/OE.20.022813.

(38)  Wu, S.; Yan, G.; Lian, Z.; Chen, X.; Zhou, B.; He, S. An Open-Cavity Fabry-Perot Interferometer with PVA Coating for Simultaneous Measurement of Relative Humidity and Temperature. *Sensors and Actuators B: Chemical* **2016**, *225*, 50–56. https://doi.org/10.1016/j.snb.2015.11.015.

(39)  Mileńko, K.; Hu, D. J. J.; Shum, P. P.; Zhang, T.; Lim, J. L.; Wang, Y.; Woliński, T. R.; Wei, H.; Tong, W. Photonic Crystal Fiber Tip Interferometer for Refractive Index Sensing. *Optics Letters* **2012**, *37* (8), 1373–1375. https://doi.org/10.1364/OL.37.001373.

(40)  Ferreira, M. S.; Roriz, P.; Bierlich, J.; Kobelke, J.; Wondraczek, K.; Aichele, C.; Schuster, K.; Santos, J. L.; Frazão, O. Fabry-Perot Cavity Based on Silica Tube for Strain Sensing at High Temperatures. *Optics Express* **2015**, *23* (12), 16063–16070. https://doi.org/10.1364/OE.23.016063.

## Tables

| Method | Sensitivity | Tolerance |
| --- | --- | --- |
| **Fiber-optic temperature sensor** [18] | 84.6 pm/°C | |
| **Optical fiber tip-based concave-shaped cavity** [25] | 12.26 pm/°C | 1000 °C |
| **Sapphire Fiber** [26] | 32.5 pm/°C | 1550 °C |
| **Based on Vernier Effect** [36] | 654 pm/°C | 120 °C |
| **Fiber in-line Fabry-Perot interferometer cavity** [37] | 4.3 pm/°C | 26.4–100 °C |
| **Fiber open-cavity Fabry-Perot interferometer with polyvinyl alcohol coating** [38] | −6.14 pm/°C | 10–50 °C |

| Method | Sensitivity | Tolerance |
|---|---|---|
| The photonic crystal fiber tip ended with a solid silica-sphere [39] | 10 pm/°C | 20–100 °C |
| Silica tube with 14 µm of cladding [40] | 0. 75 pm/°C | 22–750 °C |
| Our realization | 180 pm/°C | 20-150 °C |
| Fiber-optic temperature sensor [18] | 84.6 pm/°C | |

## Data Accessibility

The datasets supporting this article have been uploaded as part of the Supplementary Material

## Authors' Contributions

Hani Barhum: investigation, acquisition of data, writing—original draft, writing—review and editing;
Muhammad A. Atrash: investigation, writing—original draft, writing—review and editing;
Inga Brice: investigation, writing—original draft, writing—review and editing;
Toms Salgals: investigation, writing—original draft, writing—review and editing;
Madhat Matar: investigation, writing—original draft, writing—review and editing;
Mariam Amer: investigation, writing—original draft, writing—review and editing;
Ziad Abdeen: supervision, writing—original draft, writing—review and editing
Janis Alnis : supervision, writing—original draft, writing—review and editing
Vjaceslavs Bobrovs: supervision, writing—original draft, writing—review and editing
Pavel Ginzburg: supervision, writing—original draft, writing—review and editing

All authors gave final approval for publication and agreed to be held accountable for the work performed therein.

## Competing Interests

*We have no competing interests.*

# Figure and table captions

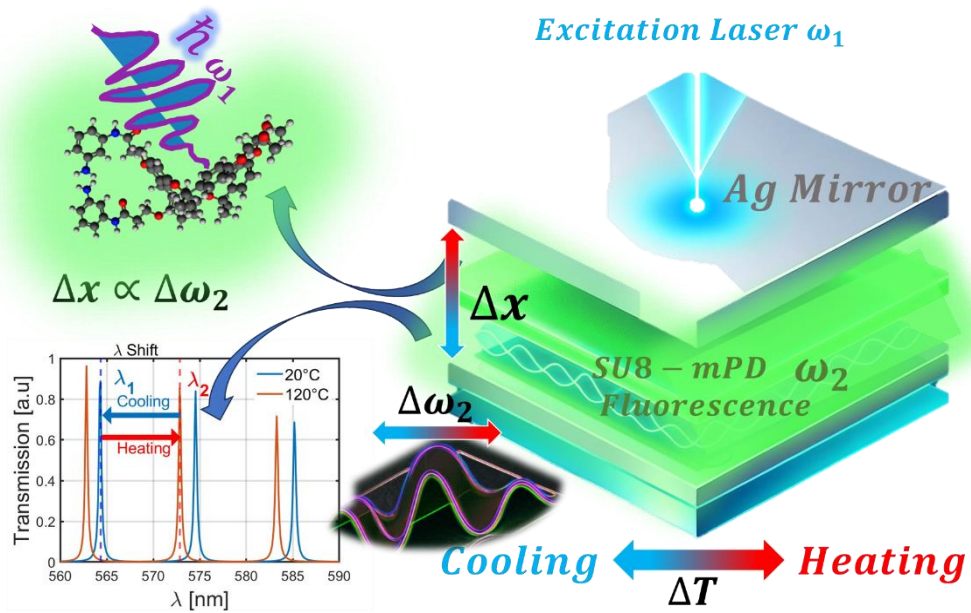

*Figure 6: Schematics of FP-based temperature sensor, encompassing two mirrors and a thermoresponsive conjugated fluorescent polymer. The structure is pumped at $\omega_1$ and emits light at a temperature-dependent wavelength. The measurement is the shift of the emission peak, i.e., '$\Delta\omega_2$', which depends on the distance between the mirrors. Thermoresponsive polymer experiences contraction or expansion '$\Delta x$', depending on the temperature change, which is monitored by '$\Delta\omega_2$'.*

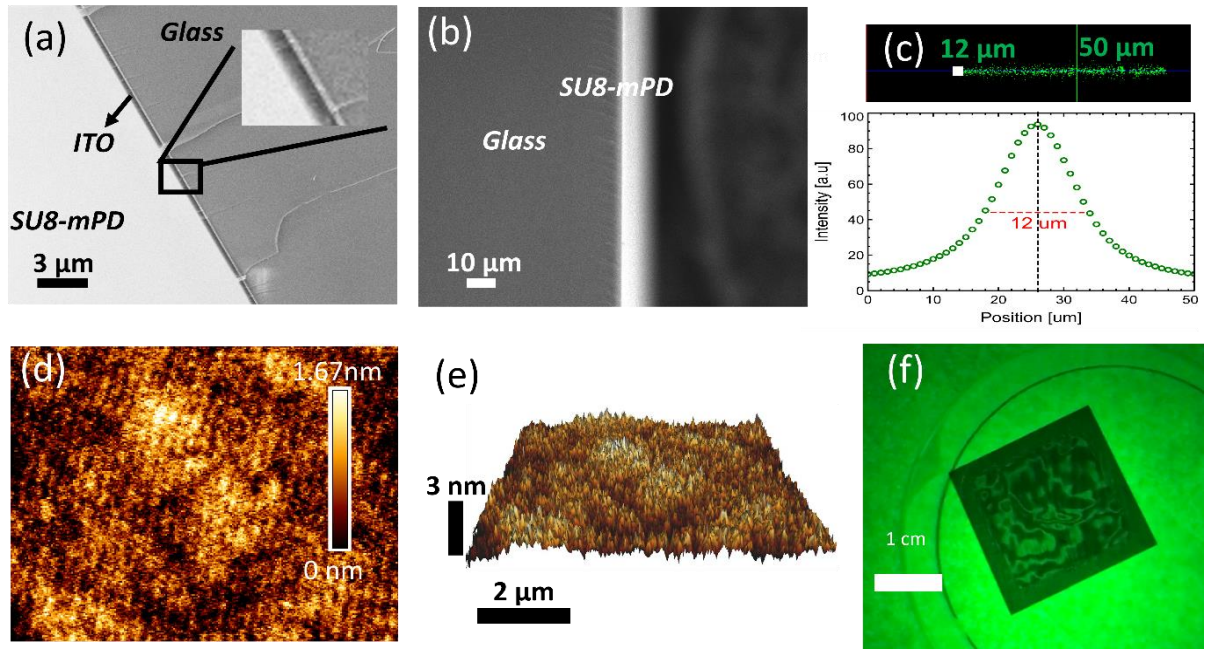

Figure 7: Characterization of SU8-mPD conjugated thin films. (a)-(b) Cross-section SEM images of SU8 films on a glass coverslip, (c) Cross-section of the fluorescent film as observed in the confocal microscope (488 nm pump). The fluorescent profile is fitter with Gaussian and the red arrow indicates its 12 μm width. (d)-(e) High-resolution AFM image of the surface roughness in 2D and 3D layouts, respectively. (f) Image of the chip under 532nm LED showing interference patterns.

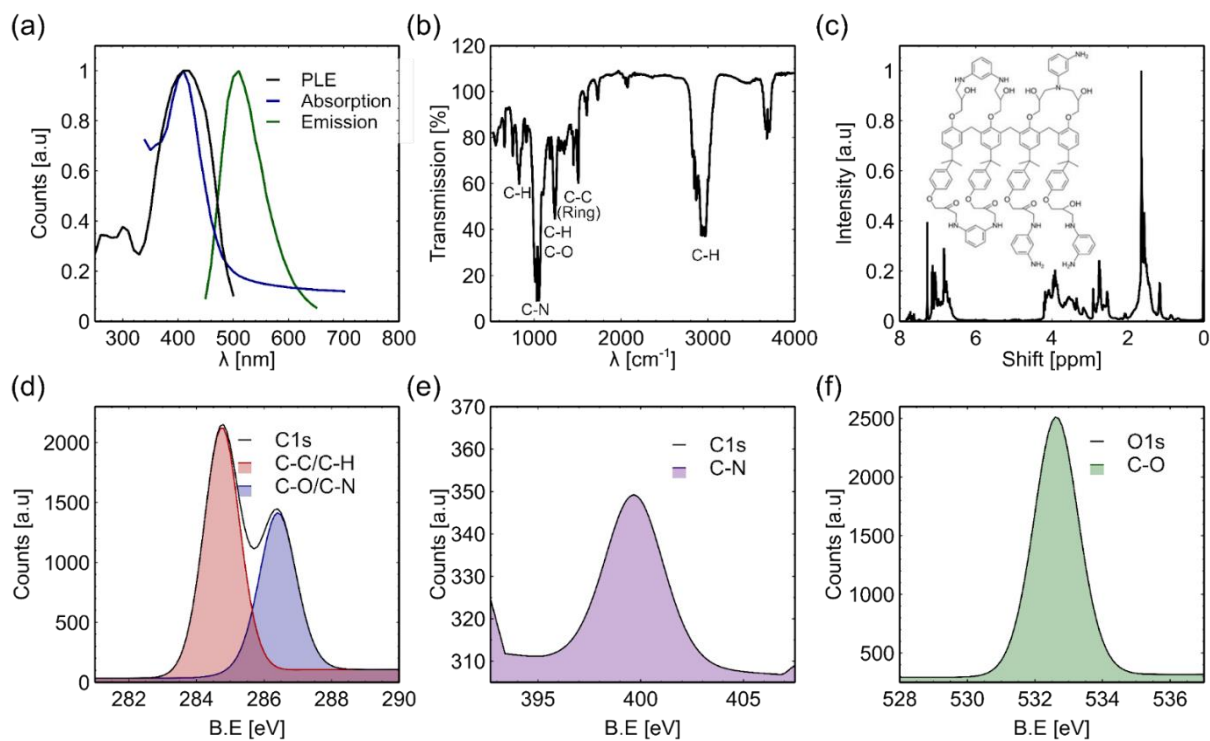

Figure 8: Optical and structural properties of SU8-mPd layer. (a) PLE (observed at 530 nm), absorption, and emission (excitation at 420nm). (b) FTIR spectrum reveals the presence of C-H, C-C, C-N, and C-O bonds. (c) NMR shift - three regions at 1.6, 3.5, and 7 ppm. The suggested polymer structure appears as an inset. (d)-(f) XPS data show the presence of carbon(C(1s)), nitrogen (N(1s)) and oxygen (O(1s)), with an atomic weight of 76 %, 1.3 %, and 22.7 %, respectively.

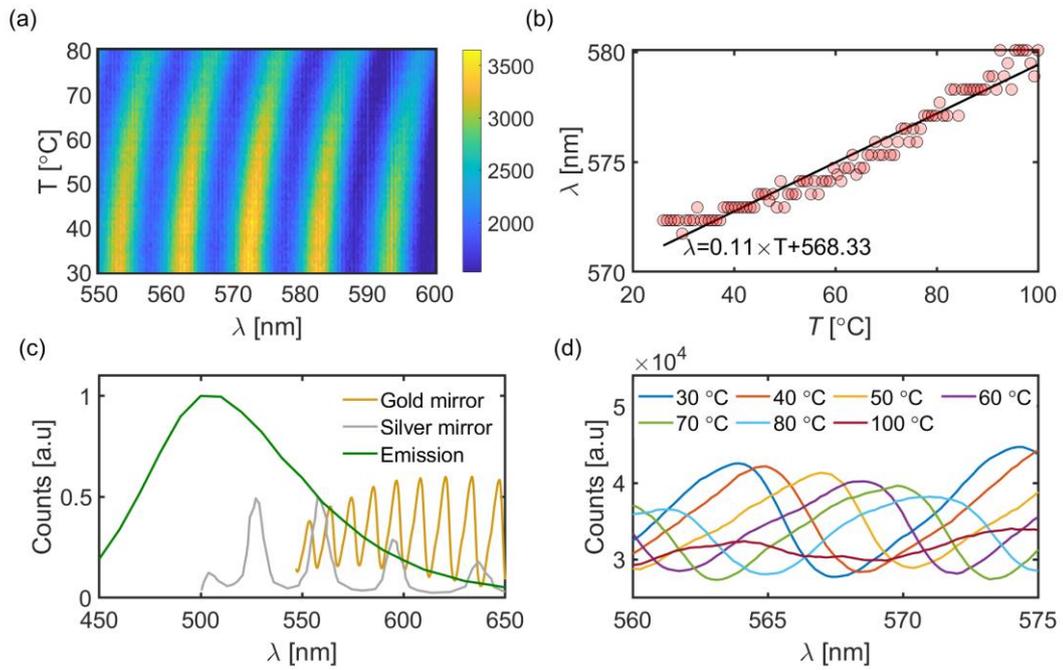

Figure 9: (a) Temperature-dependent evolution of the FP-SU8-mPD transmission. (b) Transmission peak evolution with temperature. The analysis is done on the branch, indicated with a thin black line in panel (a). (c) Green curve - emission spectrum of SU8-mPD, yellow – 10 µm-thick FP with gold mirrors transmission, and gray - 5 µm-thick FP with silver mirrors transmission. (d) Zoomed peak shift of transmission spectra at different temperatures (gold FP).

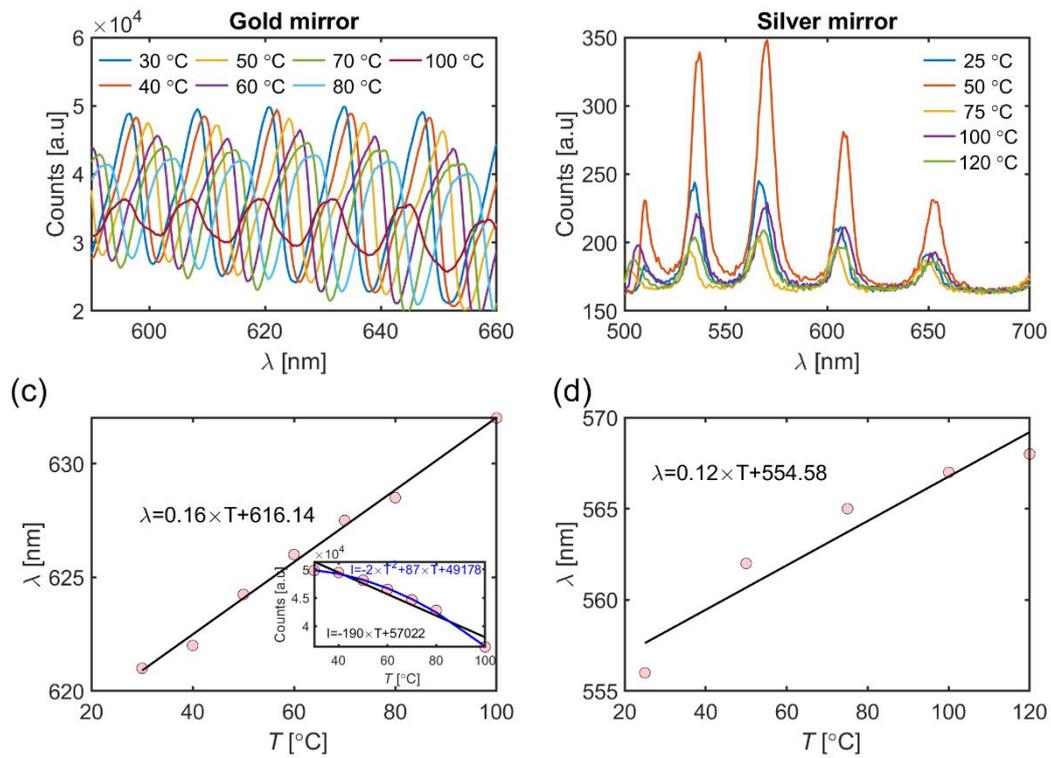

*Figure 10. Photoluminescent intensity [a.u.] at different temperatures for Fabry-Perot devices with (a) gold and (b) silver mirrors. SU8-mPD layer thicknesses are 10µm and 5µm, respectively. (c) Resonance wavelength as a function of temperature (for gold mirror) and the corresponding linear fit. Inset - photoluminescence intensity (of the peak) as a function of temperature and the first and second-order polynomial fits. (d) Resonance wavelength as a function of temperature for silver mirror device.*

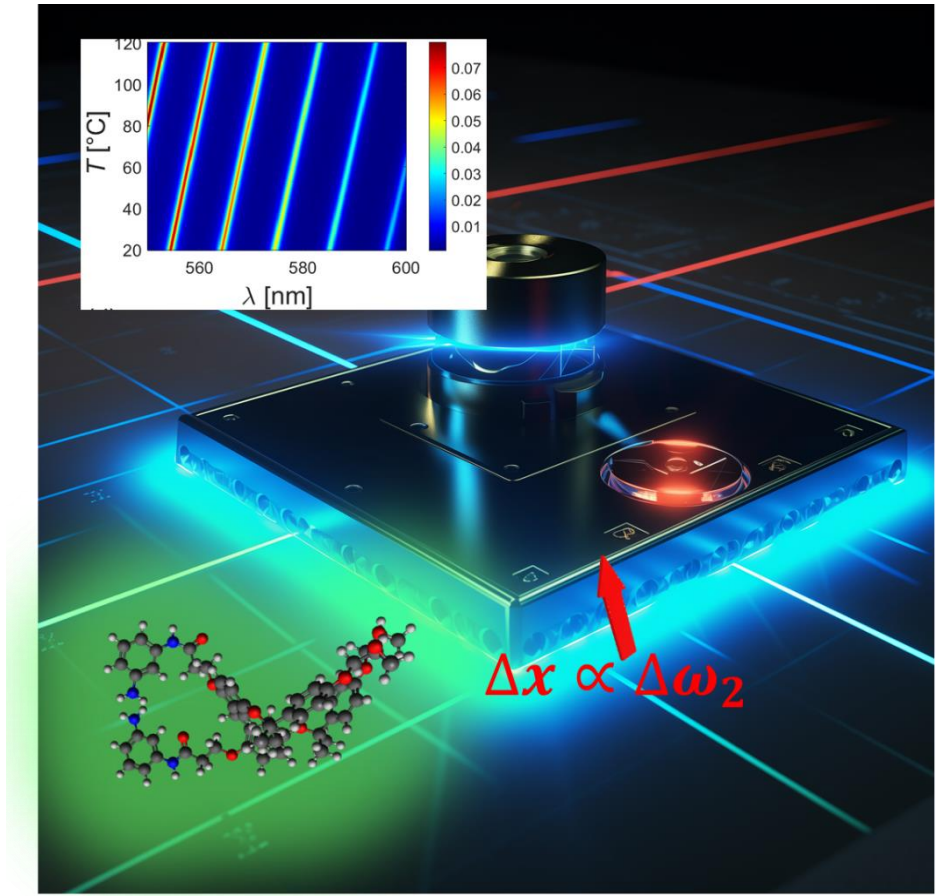

**TOC.** SU8-mPD Fabry-Perot fluorescent sensor. The optically pumped structure emits light at temperature-dependent wavelength.  Thermal responsivity is obtained via the SU8-mPD layer expansion.

# Supplementary material

# SU-8 meta phenylenediamine conjugated thin film for temperature sensing

**Hani Barhum[1,2,†,\*], Muhammad A. Atrash[1,2], Inga Brice[3], Toms Salgals[4], Madhat Matar[2], Mariam Amer[1,2], Ziad Abdeen[5,6], Janis Alnis[3], Vjaceslavs Bobrovs[4,] Abdul Muhsen Abdeen[5,7], and Pavel Ginzburg[1]**

*[1]Department of Electrical Engineering, Tel Aviv University, Ramat Aviv, Tel Aviv 69978, Israel.*
*[2]Triangle Regional Research and Development Center, Kfar Qara' 3007500, Israel.*
*[3]Institute of Atomic Physics and Spectroscopy, University of Latvia, Jelgavas Street 3, 1004 Riga, Latvia.*
*[4]Institute of Telecommunications, Riga Technical University, 12 Azenes Street, 1048 Riga, Latvia.*
*[5]Al-Quds Public Health Society, Jerusalem, Palestine*
*[6]Al-Quds Nutrition and Health Research Institute, Al-Quds University, East Jerusalem, Palestine*
*[7]Marshall University John Marshall Dr, Huntington, WV 25755, United States*
*HB, https://orcid.org/0000-0003-0214-0288 ; MAA, https://orcid.org/0000-0002-2500-0811*

## S1. Measurement of sample thickness

The layer thickness is the critical parameter, which will be extracted from the FP cavity transmission spectrum. The white light transmission (experimental data) appears in Figure S1(a). Wavelengths below 500nm are absorbed by the layer in the correspondence with Figure 2(a). The typical high-Q transmission peaks are observed at longer wavelengths. Figure S1(b) shows the Free Spectral Range (FSR) (the frequency interval between successive transmission peaks) as the function of wavelength. FSR can be calculated as $\nu = \frac{c}{2n_g L}$, where $L$ is the layer thickness and $n_g$ is the group index. Figure S1(b) demonstrates a relatively flat FSR, thus allowing to neglect of the group index dispersion, especially in cases where a 100nm fluorescent source is used (as in the rest of the experimental data). The experimental fit suggests $L = 10 \mu m$ $(n_g = 1.56)$, corresponding to SEM images in Figure 1.

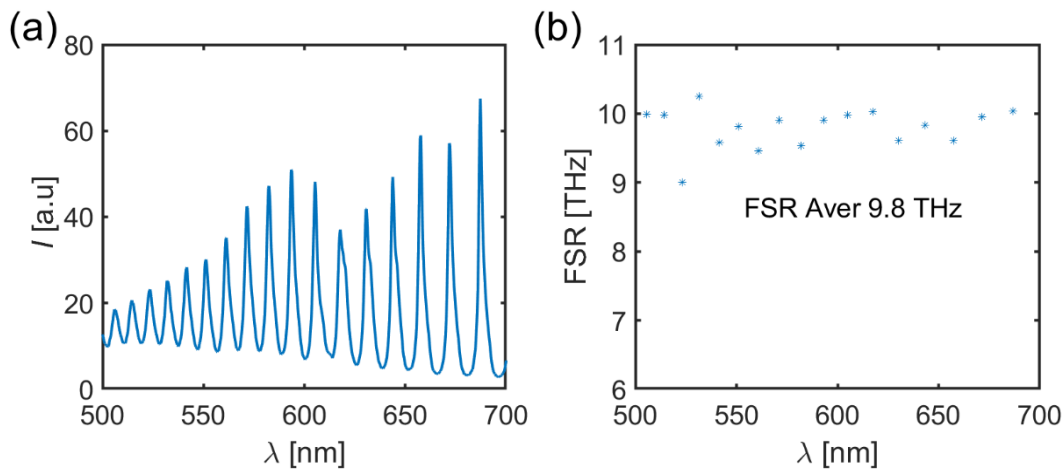

*Figure S1: Free Spectral Range (FSR) analysis to FP Cavity Thickness. (a) Transmission spectrum, (b) FSR as the function of the wavelength.*

## S2. Fabry-Pérot resonator response to temperature change – numerical analysis

The FP resonator comprises an SU8 layer sandwiched between two parallel metal mirrors, which can be either silver or gold. The resonator is illuminated by a plane wave at normal incidence for the sake of simplicity. The field amplitude and the frequency were scanned to fit a Gaussian shape, corresponding to a fluorescent spectrum of SU8-mPD, which is characterized by a peak at 530 nm and a Full Width at Half Maximum (FWHM) of 100 nm. This step was applied to fit the experimental data, where the generated finite bandwidth fluorescence is superimposed in the FP transfer function. The simulation spans a wavelength range from 550 nm to 600 nm and a temperature range from 20°C to 120°C. The SU8 layer's initial thickness, its coefficient of thermal expansion, the refractive indices of the SU8 layer, and the mirror materials (silver and gold) are crucial parameters, that govern the transmission properties.

The numerical results are visualized through 2D colormaps and 1D plots, showing the transmission intensity across the specified wavelength and temperature ranges (Figure S2). The FP resonance condition is $\lambda = 2nd/m$ where $\lambda$ is the resonant wavelength, $n$ is the refractive index, $d$ is the cavity length, and $m$ is an integer. The peak positions are slanted to the longer wavelengths with the temperature increase. This is an obvious consequence of the thickness ($d$) expansion. The temperature dependence of the layer thickness was retrieved from measurements and appears in Figure 4. Furthermore, gold, being a more lossy material than silver, broadens the FP transmission peaks and also reduces the peak value. In this numerical analysis, gold dielectric indexes were adopted. All the parameters for the analysis are summarized in Table ST1.

**Table ST1. Parameters for FP transmission**

| Parameter | Symbol | Equation |
|---|---|---|
| Wavelength Range | $\lambda$ | 550 - 600 nm, Gaussian, 530 mean, 100nm SVD |
| Temperature | T | (20°C,120°C,5) |
| Initial Thickness of SU8 | $L_0$ | $L_0$=10µm |
| Coefficient of SU8 Thermal Expansion | $\alpha$ | $\alpha$=152×10$^{-6}$[K$^{-1}$] |
| Refractive Index of SU8 | $n_{SU8}$ | $n_{SU8}$= 1.58 |
| Mirrors Reflectivity | R | Silver and Gold Layers, 30 nm thickness, refractive index from |

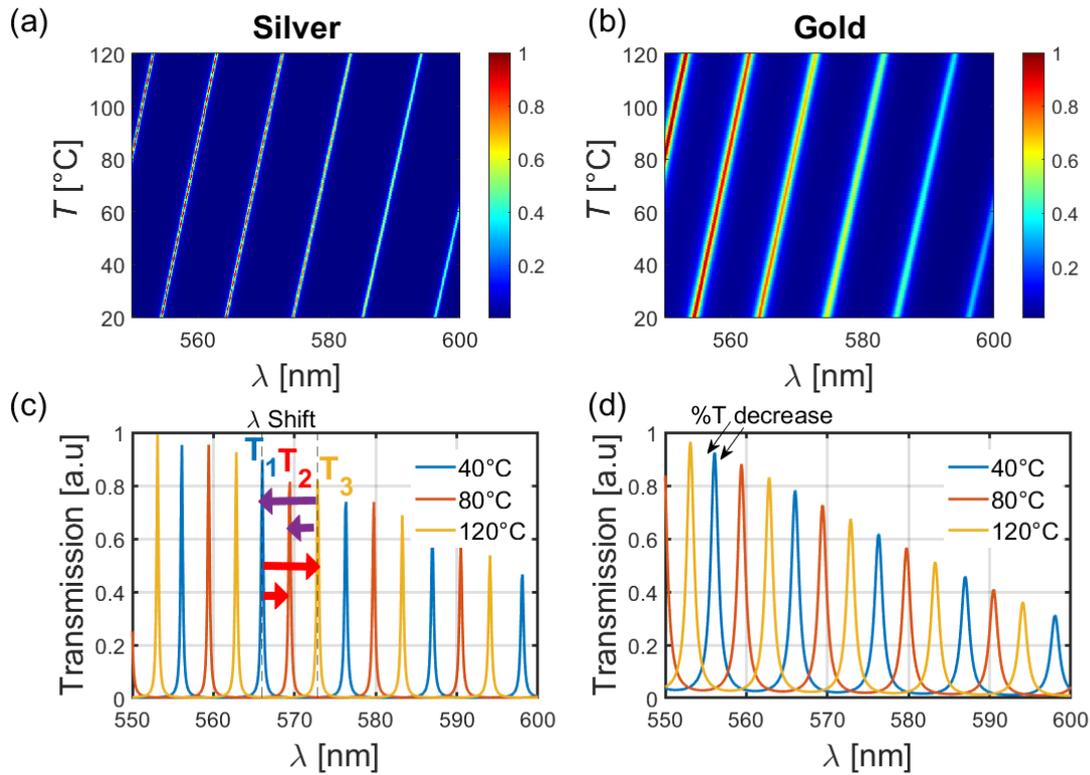

*Figure S2: FP transmission of a light source with Gaussian spectrum (550mean, SVD 100nm). (a), (b) colormaps demonstrating the transmission spectra as a function of temperature for silver and gold mirrors, correspondingly. (c), (d) slices through the colormaps, demonstrating the evolution of the peak for different temperatures (in legends).*

## S3. Fluorescence of SU8-mPD Fabry-Perot

The spectral evolution of the device, undergoing both heating and cooling cycles, appears in Figure S3 (and b). The experiments were obtained from different spots on the sample and, thus, are not in scale with each other. The measurements were not done below 500 nm due to the strong light re-absorption. Figure S3 (c and d) is the evolution of the emission peaks. The shift between heating and cooling cycles does not come from the hysteresis effect but rather the result of the sample shift during the experiment. The long-range variation of thickness (Figure 2(f)) is responsible for the effect.

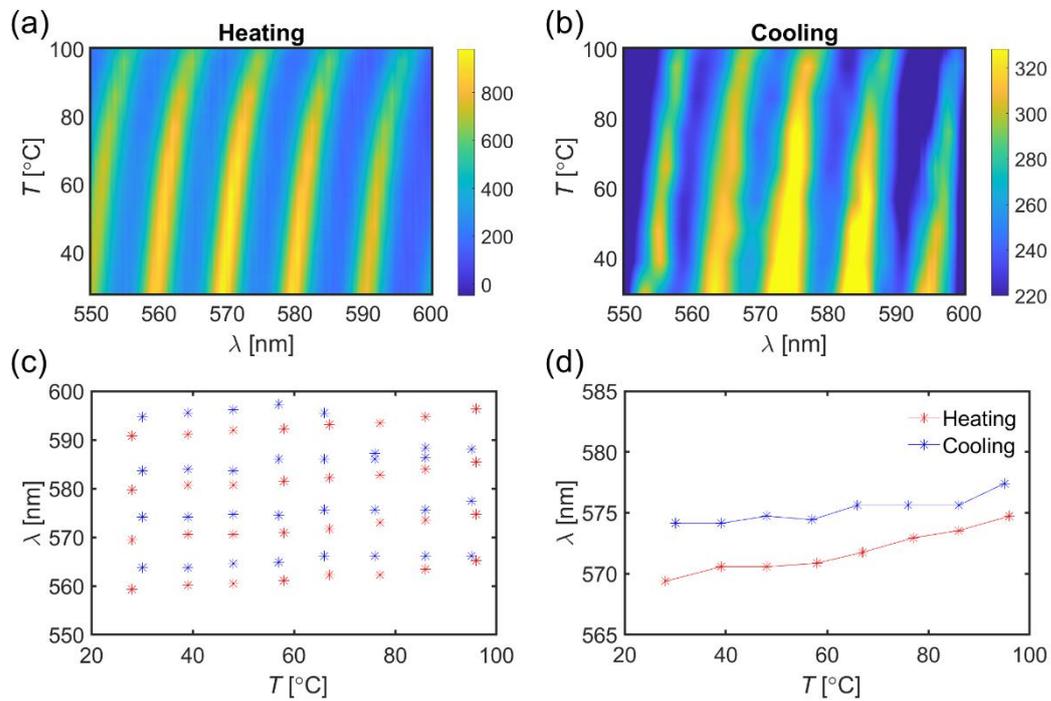

*Figure S3: Temperature-dependent photoluminescence (PL) properties of the device. (a, b) – colormaps, demonstrating the fluorescent signal spectra as a function of temperature and wavelength for (a) heating and (b) cooling cycles. (c) resonance shift of emission peaks as a function of temperature. (d) Enlarged data for one of the peak evolutions.*